# Case study of Innovative Teaching Practices and their Impact for Electrical Engineering Courses during COVID-19 Pandemic

Amith Khandakar*[1], Muhammad E. H. Chowdhury[1], Md. Saifuddin Khalid[2], Nizar Zorba[1]

[1]Department of Electrical Engineering, Qatar University, Doha-2713, Qatar;amitk@qu.edu.qa (AK), mchowdhury@qu.edu.qa (MEHC), nizarz@qu.edu.qa (NZ)

[2] Department of Applied Mathematics and Computer Science, Technical University of Denmark; skhalid@dtu.dk (MSK)

**Abstract**

Due to the COVID-19 pandemic, there was an urgent need to move to online teaching and develop innovations to guarantee the Student Learning Outcomes (SLOs) are being fulfilled. The contributions of this paper are two-fold: the effects of an experimented teaching strategy, i.e. multi-course project-based learning (MPL) approach, are presented followed with online assessment techniques investigation for senior level electrical engineering (EE) courses at Qatar University. The course project of the senior course was designed in such a way that it helps in simultaneously attaining the objectives of the senior and capstone courses, that the students were taking at the same time. It is known that the MPL approach enhances the critical thinking capacity of students which is also a major outcome of Education for Sustainable Development (ESD). The developed project ensures the fulfillment of a series of SLOs, that are concentrated on soft engineering and project management skills. The difficulties of adopting the MPL method for the senior level courses are in aligning the project towards fulfilling the learning outcomes of every individual course. The study also provides the students feedback on online assessment techniques incorporated with the MPL, due to online teaching during COVID-19 pandemic. In order to provide a benchmark and to highlight the obtained results, the innovative teaching approaches were compared to conventional methods taught on the same senior course in a previous semester. Based on the feedback from teachers and students from previously conducted case study it was believed that the MPL approach would support the students. With the statistical analysis (Chi-square, two-tailed T statistics and hypothesis testing using z-test) it can be concluded that the MPL and online assessment actually help to achieve better attainment of the SLOs, even during a pandemic situation.

## 1. Introduction

To develop a sustainable, environment-friendly, and conscious society, it is advisable to use technology, innovation, and active participation of students in the university courses. The application of the above approaches involves a wide range of abilities, like the creative resolution of problems and collective decision-making [1, 2]. The need of engineering gradutes with skills that can help in sustainable development , which is also known as Education for Sustainable Development (ESD), needs the traditional curriculum approaches to be changed [3] and several creative instructional methods need to be implemented making students active observers and prepared with ESD skill sets, which are highly demanded in the current times [4-9]. Students are encouraged to appreciate the learning effects and its applicability in overcoming challenges encountered in everyday life [3, 4]. In a recent publication [10], the authors have highlighted an improvement in the learning outcomes as a result of inclusion of ESD in the curriculum in the form of Multi course Project Based Learning, which is used as a guideline for the study discussed in this article. In their previous work [10], they stressed the applicability of MPL, i.e. a project which was used in the assessment of different courses, as the students could not create a successful working prototype using the ideas provided in each course separately. Moreover, due to the work-load of separate projects in different courses, it does not motivate the students to produce quality output. A multi-course, project-based learning (MPL) experience was shared in

[11], where MPL was utilized in a software development assignment. In a recent work, researchers stressed the notion that realistic multi-course lessons in the form of a project would help to encourage students in understanding and applying their technical expertise to an engineering project [12]. Another work used a multi-semester initiative [13] to demonstrate how capability set deficiencies in traditional curricula were resolved using the proposed approach of having a project span over many semesters. None of the works listed, however, included information about how the projects were planned and executed and did not include a comprehensive comparative study in terms of efficacy. This paper gives a detailed overview of how, in a series of steps, the authors introduced a multi-course project (Figure 1). In addition to modifying the standard curriculum to integrate ESD, one of the main reasons for this case study was to affirm the success of a multi-course initiative in achieving student learning outcomes and promoting project-based learning (PBL). PBL aims to improve study methods while studying real-world problems and is appealing to learners as well [14]. A similar MPL was implemented by the authors in [10], where they have used two junior level undergraduate Electrical Engineering courses and have shown that such an implementation is appreciated by the students and have also successfully achieved the Student Learning Outcomes (SLOs) for the courses involved. The authors have also shown a step-by-step approach how such an implementation can be done for undergraduate level university courses. This approach could be further verified in the new education paradigm ,i.e. online learning, due to the pandemic.

The coronavirus disease 2019 (COVID-19) outbreak has been declared a global pandemic by the World Health Organization (WHO) and education at all levels has significantly been affected and much research is being carried out on innovative teaching methods to reduce the loss. Many research articles were discussing the approaches adopted to tackle the hindrances caused by the COVID-19 pandemic to medical studies [15-19]. All the research works have talked about the flipped classroom paradigm, online practice questions, teleconferencing instead of in-person seminars, engaging residents in telemedicine clinics, procedural modeling, and even the use of video surgeries were encouraged as a creative alternative. Although there is no replacement for realistic learning for direct patient care, these can be ways to minimize the limitation of learning during this period. In this extraordinary scenario, innovative solutions through technology will help to fill the educational gap for the surgical residents. Wei Bao in [20] has focused on a case analysis in online education at Peking University and introduced six unique instructional techniques to outline existing online education experience for university instructors. The study concludes with five high-impact online learning principles: (a) high relevance between online teaching design and student learning, (b) efficient provision of online teaching knowledge, (c) adequate support given to students by faculty and teaching assistants, (d) high-quality engagement to expand the scope and depth of learning for students, and (e) contingency plan to deal with unforeseen online education network accidents. To the best of the authors, there is still some case study missing on the innovative teaching strategies adopted by engineering courses for retaining the education for sustainable development even during the pandemic crisis. This is where our paper contributes to the field of knowledge by showcasing a detailed framework on the use of the MPL and online teaching approach for ESD during a pandemic situation. The paper also highlightes the implication of such an approach with the help of student feedback and also student performance.

To test the programming and problem-solving abilities of undergraduate Electrical Engineering (EE) students at Qatar University, two senior-level courses were used in this study. The first one is an elective course, entitled 'Wireless Network and Application' (ELEC 472), and the second one is the 'Senior Design Project II' (ELEC 498), a capstone course. To integrate ESD in the courses and to meet the criteria set by the industry, and to raise the level of participation of students in problem-solving, and independent study, the curriculum as well as the instructional approach of the elective course have been changed. The students were asked to focus on a multi-course project and the instructions were structured to enable collaboration on a project involving the knowledge and skills learned in both courses. The project-grading rubric was also structured to fulfil the requirements for the learning outcomes of both of the courses. The study was conducted during the Spring semester (January to May 2020).

This article is arranged as follows. Section 2 describes the multi-course project-based learning approach applied in this study. Section 3 presents the stages of the execution of this study customized for a senior level course. Section 4 presents the results of the analysis and an assessment of the impact of MPL study in terms of how well the students have attained the learning outcomes. Finally, Section 5 discusses the findings and recommendations for the changes based on input from the instructors and students.

**2. Fundamental block/concept behind the study**

The project-based learning (PBL) concept, fundamental block of MPL, helps the students to implement the project with self-learning and minimal mentoring [10, 21]. PBL depends on an involved, interconnected, and positive learning process, informed by social and contextual influences, which is necessary for learners to develop skills for an improved sustainable future [22-24]. The PBL methodology should provide the following features: student-centered learning in limited students group, instructor as a tutor or guide, and challenges proposed to help the student self- learn the course-related advanced knowledge that can be used to overcome them [9, 24]. The successful introduction of PBL in engineering education at the University of Aalborg, Denmark can always be used as a reference [24-26].

In a multi-course project, explicit project rules/instructions stating the deliverables and detailed rubric to grade are key parameters [10, 21]. The instructions will prevent any misunderstanding among the students about how to present the project and illustrate the specifications of each course. It is also important that the project-grading rubric should be following the learning outcomes of the involved courses. The MPL requires specific and concise guidance, including coordination and adaptation to the previous experience of the students [27]. MPL believes in engaging in groups to solve challenges together with their creative thinking, where the teacher is just a facilitator in collective learning [22-24, 28]. These methods help students to be prepared for the competitive market that is always seeking people with skills for a sustainable solution and also can communicate it intuitively to the audience [29]. Studies involving surveys on the effects of PBL on ESD teaching in European institutions of higher education have been conducted in [30] and in technical university in Malaysia in [31] with positive results provided confidence to the study. In Germany and Vietnam, similar studies were also performed in [32] and [33], respectively. As stated earlier , all the studies did not provide the step by step approach of how such Multi Course project can be implemented and its effect analysed. This study is trying to fulfill the gap.

## 3. Research Method

This study redesigns and tests a multi-course project-based learning case in four phases: a review phase, a design phase, an implementation phase, and an analysis phase (Figure 1).

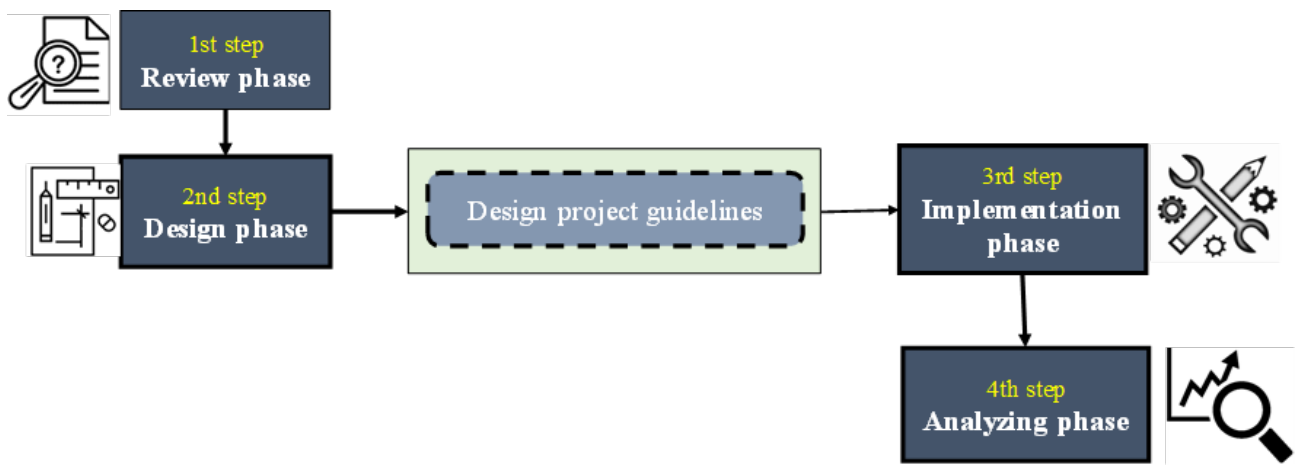

Figure 1. Important phases of MPL.

The methodological approach was discussed in detail in the authors' previous work [10]. This can be considered similar to the process groups of a project (according to Project Management Book of Knowledge PMBOK)- Initiation, Planning, Execution, Monitoring and Control and Closing, with difference being that Monitoring and Controlling and Closing phase is combined as analyzing phase. As in any project, the planning phase is the most crucial and it is the review and design phase when it is done. The first phase of action preparation or analysis entails the teacher planning the activities to be performed in the course, in the form of relating the learning methods of the two courses, one elective and the other one is a capstone course, as a common students' assignment. The concept of the project and its evaluation schedule are conducted in the second phase. In the third phase, the execution of the action plan is carried out in accordance with what was designed/planned. The fourth stage is an observation and reflection phase, in which actions are documented during the research. At the end of the analysis, the reflection process is carried out with the data collected to criticize the method and recommend the appropriate changes. In the following sections, the details ( study participants and the details of the phases) are presented.

*3.1 Study Participants*

As stated earlier, this study was conducted at the Department of Electrical Engineering, Qatar University, Doha, Qatar. In the Spring 2020 semester, the thirty-three students of the Elective Course entitled, ELEC472: Wireless Network and Applications and ELEC 499: Senior Design Project II, were involved in this study. There were 19 male students and 14 female students who participated in this study.

*3.2 Review Phase*

The Electrical Engineering Department of Qatar University is an Accreditation Board for Engineering and Technology (ABET) accredited unit of higher education. All the courses have the Course Learning Outcomes (CLOs) which have to be linked to the ABET student learning outcomes. Thus, the course instructors evaluated the learning outcomes of the senior course (ELEC 472) and tested if there was any chance of designing a project that could be used to test courses' learning outcomes and at the same time help in fulfilling some of the course learning outcomes of the capstone course (ELEC 499). The mapping of the corresponding CLOs with the SLO's can be seen in Figures 2 and 3.

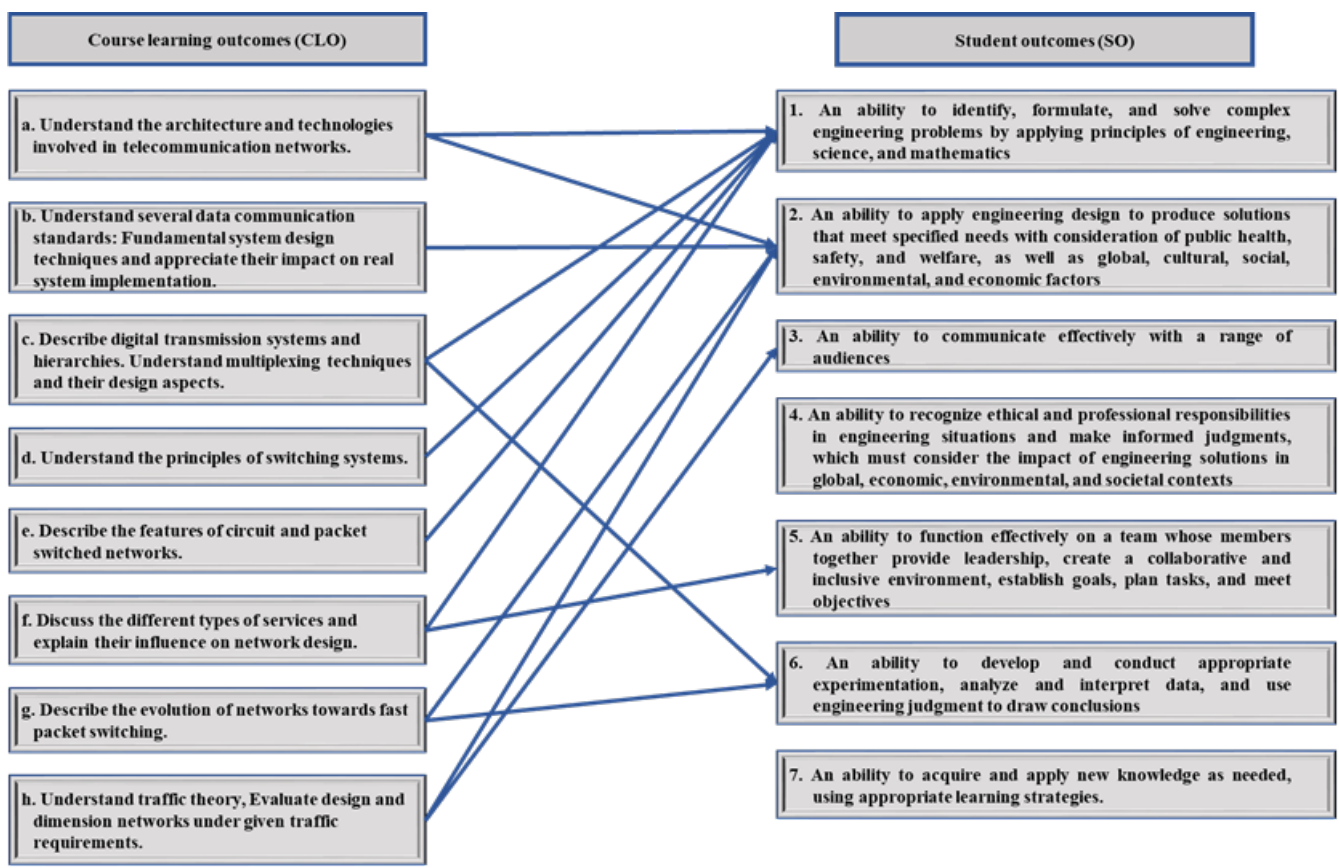

**Figure 2.** CLO and SLO mapping of ELEC 472

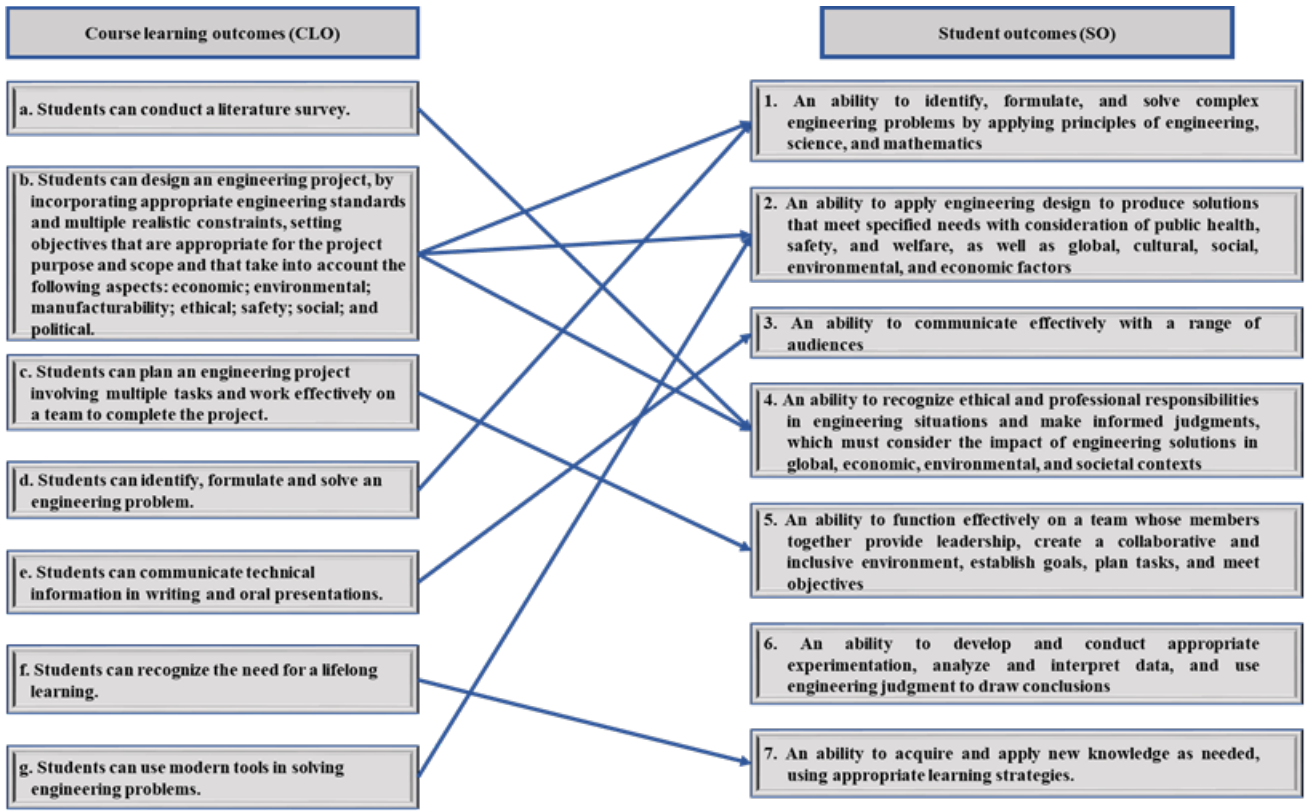

**Figure 3.** CLO and SLO mapping of ELEC 499

A MPL was finalized as an assessment method to fulfill most of the course learning outcomes for the two courses (ELEC 472 and 499) after close analysis of CLOs with the SLOs for the courses. A project for the design concept was allocated to analyze the cumulative CLOs for these courses. The project was to develop a mobile application that can be used to present the results of their capstone project (ELEC 499) using the concepts students are learning in their ELEC 472 course. The project guidelines can be found in Figure A1 in the appendix. The project was used to evaluate the CLOs "h" and "e" in ELEC 472 and ELEC 499 respectively. The CLO's help in achieving the common SLO 3. In addition to the MPL as an assessment approach, the author have also tried Online exams as assessment approach to avoid the students coming to the campus physically due to the pandemic.

*3.3 Design Phase*

Once the CLO's mapping to SLO's was decided for the course, the next step is to design both a detailed Project guidelines and the respective grading rubric. This would provide the details to the student for implementing the project for evaluation. In addition the students were also helped with training sessions with tools that can help in efficient implementation of the project. A detailed orientation was conducted to introduce the unique learning results of each course that would be measured using the MPL. The orientation included discussing the flow of the course throughout the semester and what was expected from them. The orientation is really important to have the students have confidence in the approach else they will not enjoy and learn from the process. The orientiation is always followed with the course syllabus provided to the students, which is a form of legal agreement between the students and the instructors. Also, it was verified with discussion amongst the teaching staff (Instructors and Teaching Assistants) that the MPL helped enrich the attainment of CLOs' and the analysis at the end can be used to have measurable impact. To provide realistic training on some of the topics required for the project, dedicated training sessions were organized. The trainings included lab tutorials on the use of the application that can be used by the different team members, how to manage their time effectively with the help of periodic meetings, meeting minutes, task assignments and followup. According to the course-learning outcome of the two courses, project grading rubrics and weights were defined. Sample Project guidelines (Figure A1) for ELEC 472 can be found in the Appendix.

*3.4 Implementation Phase*

This research process is involved in tracking and assessing the success of the study during the project and concludes with the fulfillment of project activities. The students are supported in executing their assignments using the comprehensive training session. Before the specific due date, the instructors have set deadlines to check the student's development to give practical insights to help them progress.

The opportunity of self-learning in projects can help in increasing creativity. It can also enhance the practical application of the concepts students are learning in the course, i.e. Mobile Application development to enhance their implementation in the capstone course. This can also help in developing a complete intuitive solution for even the examiners during the capstone course demonstration. It also helps the students in understanding more than what was discussed in the lectures through self-learning techniques. Three or four students who are working together in the capstone course form a team. With the conceptual phase of the project, this simple multi-course process begins, then verifies the structure in the simulation phase, and then tests it by integrating with their capstone prototype.

*3.5 Analysis Phase*

The design of the learning activities and students' perseptions are invested using the survey designed in the define stage and the impact is investigated based on the project grades. The participants were advised that the research was to see the students' views on the efficacy of multi-course projects in the Electrical Engineering courses described above. The question was specifically crafted so that no redundant questions were asked and the experience of the authors was used to prepare surveys that gather the information that can be tested for relevant conclusions [34]. Strict steps were taken to protect the privacy of the participants and the secrecy of answers by preventing the identification of the participants. Besides, the cumulative evaluation and disclosure of the data prohibited participants from being identified. The survey questionnaire was conducted to acquire students' feedback and questionnaire reviews as seen in Table 1. From the prior experience of the authors, who were the instructors of these courses, and feedback from students who took the course earlier, a successful and effective learning experience from the project approach was expected. Gender-based responses have been analyzed to further verify if the approaches are typically effective in a particular student community or for different gender groups. At the end of the semester, written evaluations of students and professors, students' project scores in the previous semester, and the current semester in which the multi-course project was conducted were recorded to evaluate whether the results comply with the expectation or not. The students' multicourse project scores were used for assessment. Further many of the exams were changed to online assessment to analyze the effect of online assessment.

**Table 1.** Survey questionnaires.

| Categories | Questions Statements |
|---|---|
| Online Assessment technique | • Was the time for the online quizzes enough?<br>• Are online quizzes as convenient as paper-based quizzes?<br>• Was the format of online quizzes and feedbacks helped in a clear understanding of the concepts?<br>• Were the online video lectures convenient due to the options of stop, play, and pause? |
| MPL approach for a senior-level course and the capstone course | • Was the Mobile Application development project of the course an effective self-learning process?<br>• Was the Mobile Application development project helped in your Senior Design Project (ELEC499)?<br>• Do you think the Mobile Application development project helped in developing real-life problem-solving skills?<br>• Did the Mobile Application development project work in a group help in improving teamwork skills? |

The authors have used statistical analysis tools such as two-tailed T-test, Chi-square analysis and hypothesis testing using z-test on R software.

T-statistics are used to compare the mean values of two distinct populations. Survey findings from two classes are very useful evidence to evaluate using t-test theory since it compares and evaluates the statistical difference. The authors have used two-sample t-test analysis for the analysis in this paper, which deals with the means of the male and female responses $\mu_m$ and $\mu_f$ and finds whether or not the means are significantly different from each other by comparing the significance of t-statistics with the critical value. A t-test can only distinguish if the means are significantly different and the user decides to make a meaningful conclusion from this difference.

Each question in the survey had 5 options which were coded from 1 to 5 (1=strongly unwilling, 2=unwilling, 3=neutral, 4=willing, and 5=strongly willing respectively) and can be seen in Figure 5.

A statistic of Chi-Square ($\chi 2$) is a method [35] that tests how assumptions correspond with the evidence (or model results) currently observed.

The formula for a chi-square ($\chi 2$) statistic is

$$\chi_c^2 = \sum \frac{(O_i - E_i)^2}{E_i} \qquad (1)$$

where c is degrees of freedom, O is the observed value(s), and E is the expected value(s).

### 4. Results

A specific analysis was done to investigate the effectiveness of the approaches (MPL, Online Assessment, and Online Lecture) in the study using (i) Course overall grade in the implemented semester versus the previous semester, (ii) chi-square and t-test statistics of the survey study, and (iii) the creative-way-of-thinking indicator.

- Course Overall grade

The course grades of the students were compared for a previous semester and for the Spring 2020 semester, where the multi-course project, online assessment, and online lecturing were implemented. Course-wise, overall course grades from different semesters were compared and gender-wise grades were compared as well.

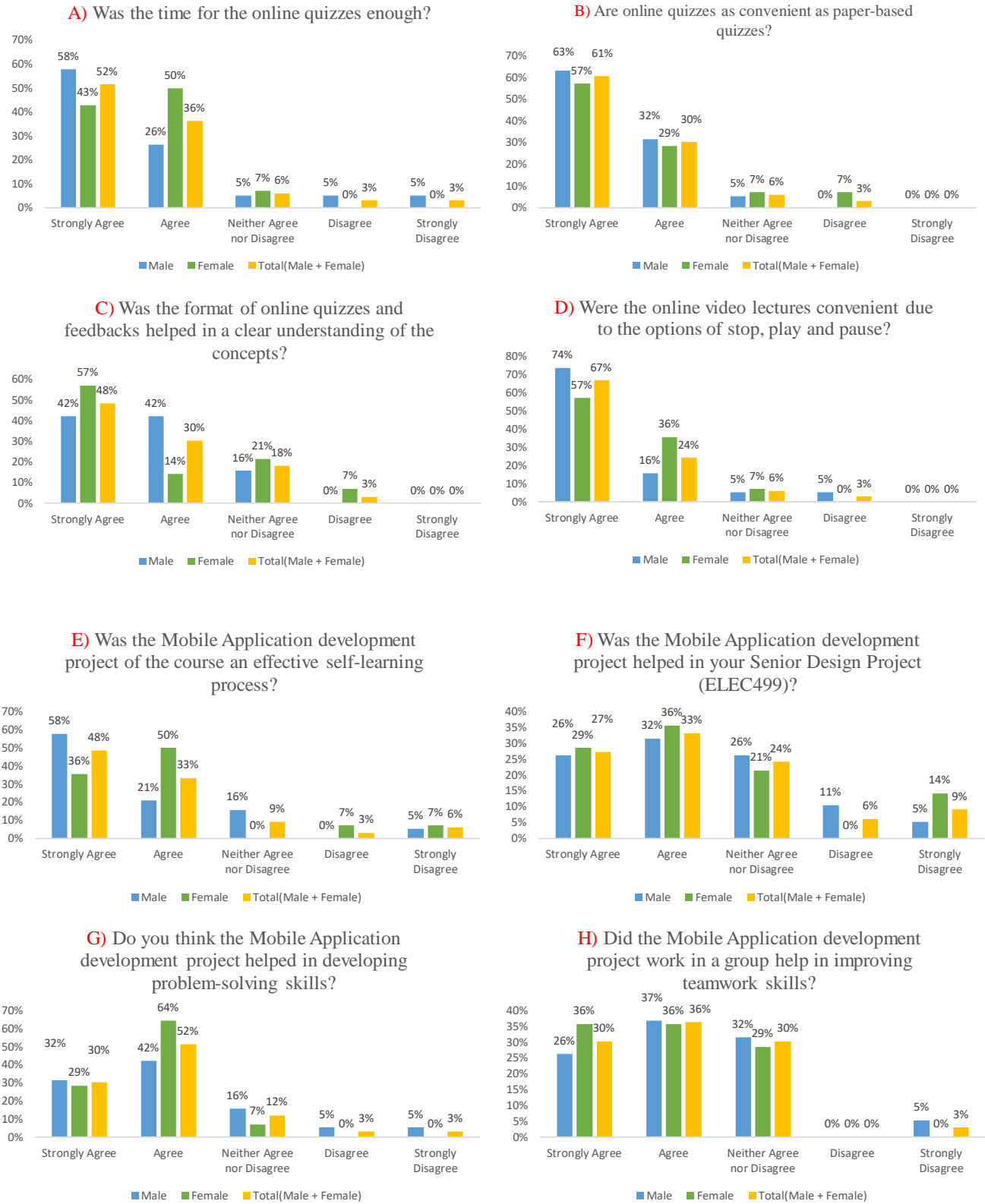

Figure 4: Summary of the student response from the survey questions

The authors have designed the survey questions in a way to get responses that can help examine the conclusions of whether the students found the MPL, online assessment, and online lecturing helpful. The authors also expected positive response on their implemented approach as it was also discussed with instructors who taught the course in previous semesters, which is also the forecast of the authors. This was tested with the use of chi-square distribution [30] as the most effective approach to test survey results [31]. Based on the results obtained from the survey and doing chi-square analysis assuming alpha level ($\alpha$) = 5 percent, the degree of freedom (c) = 4 (i.e. number of categories-1, there were five categories as shown in Figure 4). For all the questions, the authors have assumed the null hypothesis ($H_0$) as :students would respond positively (Strongly Agree or Agree). The analysis helped in accepting or rejecting the hypothesis and thus validating the assumption.

**5. Discussion on the impact of the approaches used in the study**

The impact of the approaches of the paper were analyzed in three ways :1) in terms of overall course grade from a previous semester when the same course was taught by the same instructor, 2) in t-test statistics to compare the similarity between the responses of Male and Female students, 3) chi-square statistics to compare if the survey results is as per the assumptions perceived by the author while implementing it, 4) creative thinking of the students.

*5.1. Overall Course Grade*

Figure 5 demonstrates a comparison graph between the letter grade distribution between the different semesters. It can be seen that there has been positive effect of the MPL and online assesment approach on the students' grades in the 'In-COVID semester'. A higher number of students got better grades (A, B+, and B). Moreover, it is also interesting to see that none of the students were in the lower grades such as grades D and F.

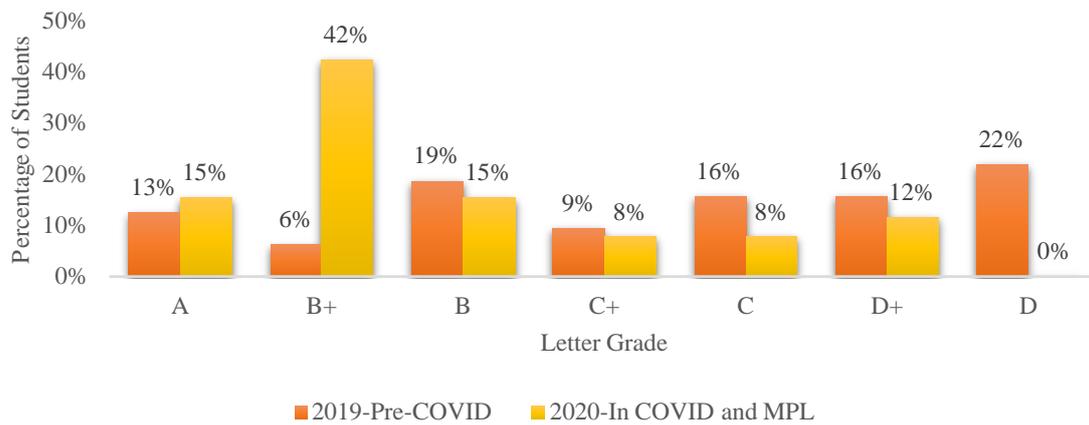

Figure 5: Comparison of the overall course grade in the two semesters

The improvement of the student perofrmance in the MPL implementation during the COVID teaching period was also confirmed using the z-test using R-software where the below hypothesis were accepted using the statistics in Figure 5.

Hypothesis ($H_0$): The probability of getting an A and B+ is larger in MLP designed course participants than among the non-MLP students.

*5.2. Analysis of the T-test Statistics*

A detailed two-sample t-test was conducted on the survey results from the Male and Female students. The responses of the students in Figure 5 were used in the two-sample t-test analysis to conclude from the student responses. The conclusions are stated in the table with the help of the t-test analysis. This will be used to suggestions and concluding remarks to make from this case study and also assess the findings of the study. Moreover, the questions were designed specifically to get feedback on the online assessment technique and the MPL approach, as shown in Table 2.

Table 2. Outcomes and conclusions from the *Two-sample t-test* on the survey answers

| Category | Questions | Two-Sample *t*-test Analysis | Conclusions Drawn |
| --- | --- | --- | --- |
| Online Assessment technique | Was the time for the online quizzes enough? | $\mu_m$ = 1.74<br>$\mu_f$ = 1.64<br>Not Significantly different | The quizzes that were covering important topics to be assessed in the course were designed well with the majority of students agreeing that the time was enough. |
| | Are online quizzes as convenient as paper-based quizzes? | $\mu_m$ = 1.42<br>$\mu_f$ = 1.64<br>Significantly Different | The response is mixed with a majority of the male have responded that the online quizzes are as convenient as written quizzes whereas some females disagreed. |
| | Was the format of online quizzes and feedbacks helped in a clear understanding of the concepts? | $\mu_m$ = 1.74<br>$\mu_f$ = 1.71<br>Not Significantly Different. | The response is really important to understand the effect of the quizzes helping to comprehend the concepts of the course. The similar positive response of both the students is really motivating. |
| | Were the online video lectures convenient due to the options of stop, play, and pause? | $\mu_m$ = 1.42<br>$\mu_f$ = 1.5<br>Significantly Different. | The response is positive with the majority of the students in favor of the videos due to the convenience but some male students disagreed with it as it provides less interaction |
| MPL approach for a senior-level course and the capstone course | Was the Mobile Application development project of the course an effective self-learning process? | $\mu_m$ = 1.74<br>$\mu_f$ = 2<br>Significantly Different. | There were mixed responses from the students through the majority of the male and female students agreed to the effective self-learning process involved in the assigned project. This feedback is important to be considered in a future implementation where more guiding materials can be provided to the students to enhance their self-learning skills. |

| Was the Mobile Application development project helped in your Senior Design Project (ELEC499)? | $\mu_m$= 2.37 $\mu_f$= 2.36 Not Significantly Different. | This was an important responsibility as it helped in understanding the effect of MPL in achieving CLO's from both courses and the students understanding it. Both the male and female had the majority of the students in favor of the MPL approach but there were some groups of students who were not impressed with the extra work involved in the project. |
|---|---|---|
| Do you think the Mobile Application development project helped in developing real-life problem-solving skills? | $\mu_m$= 2.11 $\mu_f$= 1.79 Not Significantly Different. | This was an important response as it helped in understanding the effect of MPL in improving their Problem solving skills. Both the male and female had majority of the students in favour of this aspect of implemented MPL approach. |
| Did the Mobile Application development project work in a group help in improving teamwork skills? | $\mu_m$=2.21 $\mu_f$= 1.93 Not Significantly Different. | This was an important response as it helped in understanding the effect of MPL in improving teamwork skills. Both the male and female had majority of the students in favour of the this aspect of implemented MPL approach. |

*5.3. Chi-Square Statistics*

The implementation were done with positive assumptions which were verified with statistical analysis done using chi-square analysis on the students' responses, as shown in Table 3. Overall results were similar to analysis done separately on Male and Female students (Table 3).

**Table 3.** Chi-square statistics on survey responses.

| Question Statement | Accept or Reject the positive prediction based on the Chi-Square Analysis  (chi − square  critical  value = **9.49** with Degree of Freedom = 4, α = 5%) |
|---|---|
| The time for the Online Quizzes were enough | $\chi^2_c(all\ students) = 1.87$ $< Chi - square\ critical\ value$ <br> $\chi^2_c(male\ students) = 2.25$ $< Chi - square\ critical\ value$ <br> $\chi^2_c(female\ students) = 2.80$ $< Chi - square\ critical\ value$ <br> Thus, Do Not Reject |
| The Online Quizzes are as convenient as the Paper written quizzes | $\chi^2_c(all\ students) = 4.33$ $< Chi - square\ critical\ value$ <br> $\chi^2_c(male\ students) = 3.5$ $< Chi - square\ critical\ value$ <br> $\chi^2_c(female\ students) = 1.87$ $< Chi - square\ critical\ value$ <br> Thus, Do Not Reject |
| The format of Online Quizzes and feedbacks helped in clear understanding the concepts | $\chi^2_c(all\ students) = 38.73$ $> Chi - square\ critical\ value$ <br> Thus, Reject <br> $\chi^2_c(male\ students) = 5$ $< Chi - square\ critical\ value$ <br> Thus, Do Not Reject <br> $\chi^2_c(female\ students) = 13.47$ $> Chi - square\ critical\ value$ <br> **Thus, Reject** |
| The Online Video Lectures were convenient due to the options of Stop, Play and Pause. It helped in going back to them for clarifying concepts | $\chi^2_c(all\ students) = 7.53$ $< Chi - square\ critical\ value$ <br> $\chi^2_c(male\ students) = 7.63$ $< Chi - square\ critical\ value$ <br> $\chi^2_c(female\ students) = 2.67$ $< Chi - square\ critical\ value$ <br> Thus, Do Not Reject |
| The Mobile Application Development Project of the course was an effective self-learning process | $\chi^2_c(all\ students) = 5.13$ $< Chi - square\ critical\ value$ <br> $\chi^2_c(male\ students) = 8.13$ $< Chi - square\ critical\ value$ <br> $\chi^2_c(female\ students) = 1.97$ $< Chi - square\ critical\ value$ <br> Thus, Do Not Reject |
| The Mobile Application Development Project helped in preparing for the Senior Design project (SDP 499) as it was linked with it. | $\chi^2_c(all\ students) = 53.47$ $< Chi - square\ critical\ value$ <br> Thus, Reject <br> $\chi^2_c(male\ students) = 18.63$ $< Chi - square\ critical\ value$ <br> Thus, Reject <br> $\chi^2_c(female\ students) = 6.67$ $< Chi - square\ critical\ value$ <br> Thus, Do Not Reject |

| The Mobile Application Development Project in group helped in improving real-life problem-solving skills | $\chi_c^2(all\ students) = 10.93$ $> Chi-square\ critical\ value$ Thus, Reject $\chi_c^2(male\ students) = 4.5$ $< Chi-square\ critical\ value$ Thus, Do Not Reject $\chi_c^2(female\ students) = 5.87$ $< Chi-square\ critical\ value$ Thus, Do Not Reject |
|---|---|
| The Mobile Application Development Project in groups helped in improving teamwork skills | $\chi_c^2(all\ students) = 84.27$ $> Chi-square\ critical\ value$ Thus, Reject $\chi_c^2(male\ students) = 27.75$ $> Chi-square\ critical\ value$ Thus, Reject $\chi_c^2(female\ students) = 11.17$ $> Chi-square\ critical\ value$ **Thus, Reject** |

*5.4. Students' Creative Way of Thinking*

The creativity of the students in developing the project was also explored while grading the final submissions. The clear objective provided for the project and the linking with another course helped in dedicating time effectively in preparing projects and also allowed the students to save their time to be more creative. There was a lot of scope for being creative in the project, which can be seen from some of the sample project screenshots in Figure 6.

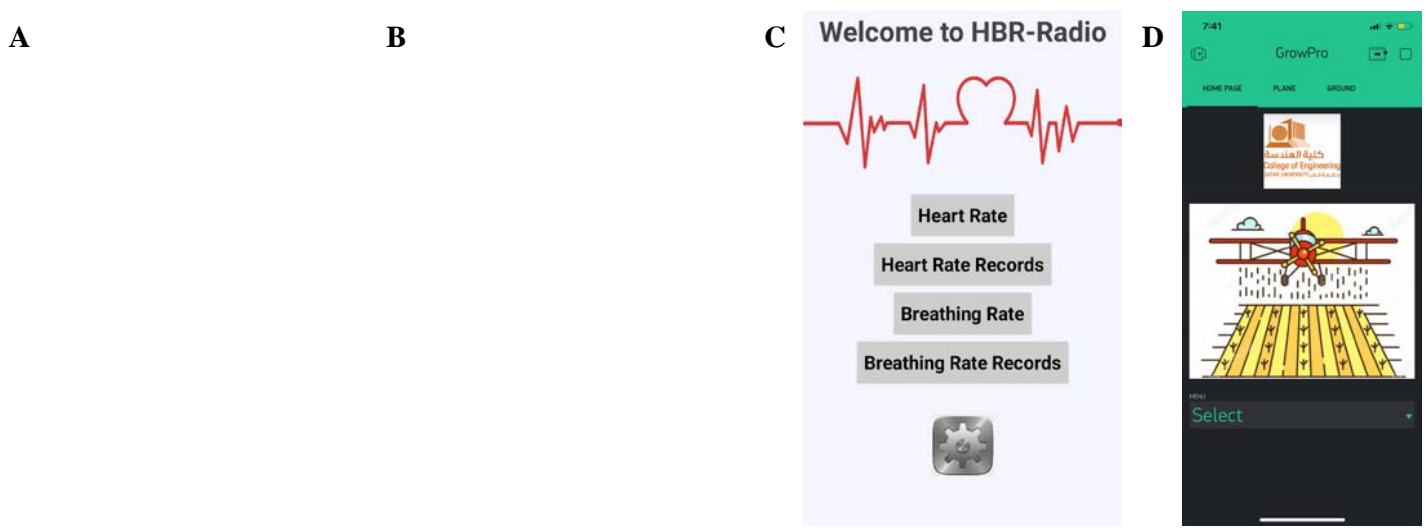

Figure 6: Screenshots of Mobile Application developed by students linking to their ELEC 499 course. A) Mobile Application to book EV charging station, B) Mobile Application to book drone cleaning services for the exterior glasses of a building, C) Mobile application to monitor the ECG of the patients, D) Mobile application to monitor the crop health and take necessary actions using drones.

**6. Conclusions**

The proposed strategy is shown to be successful in promoting PBL's core values, while some challenges are outlined, where some students were unable to develop their comprehension because they felt the learning was too difficult and needed additional tasks. Other students had troubles operating in groups or teams, by either not contributing to the group projects or by heavily contributing and upsetting other team members. The teacher must be imaginative in order to increase the motivation of such students. To enhance teacher-student relationships, teachers must be able to solve problems, especially when working with students who have low abilities, motivation, or lack concentration [10]. Furthermore, suggestions were collected on how to further to enhance the experience. The students also had concerns with interaction due to online assessment and teaching. Based on the students' responses, several modifications can be made in future studies to improve the MPL and the online teaching experience.

a) The online quizzes can be improved with the help of equations' editors and training the students on how to use them, as the students should be able to quickly write equations in their answers and then deduce the answers.

b) The feedback on the quiz should be more detailed and referring to the corresponding chapter, so that the students can use it to review for future exams and also it can be provided immediately after the test is taken. In order to do so a large pool of questions should be provided to avoid students who finish the exam earlier, sharing the questions and answer with the other students. The feedback mechanism can also help the student to understand the concepts better.

c) The online video lectures can be posted in advance for the students before the lecture, so that during the actual lecture session more interactions can be possible by letting the students ask questions.

d) The Mobile Application development project can be a more effective self-learning process with the help of more guiding materials.

The authors have found that this MPL study can add to the existing knowledge contributing by PBL studies [32-34] in the following manner:

- The step-by-step implementation of MPL for a senior and capstone course which can assist in equipping students with competencies required for sustainable development.
- The paper shed the light on how online assessment can help in improving the students' performance and learning, especially in pandemic situations.
- Some of the changes that can be made to make the methods more successful have also been outlined.
- The authors have also found a positive response for the innovations implemented in this study, verified from Table 3.

This paper discusses the effect of online assessment and MPL implementation for two senior-level courses and was designed to report the innovative teaching methods that can be useful for the pandemic situation.. Based on the students feedback, suggestions can be made that would be beneficial to improve the approach. The case study of an online course along with MPL for senior-level course can also help in adding to the existing innovative teaching practices body of knowledge.

**Author Contribution**

Experiments were designed by AK and NZ. Results were analyzed by MEHC, NZ, MSK All the authors were involved in the interpretation of data and paper writing and revision of the article.

**Funding**

NA

# Appendix

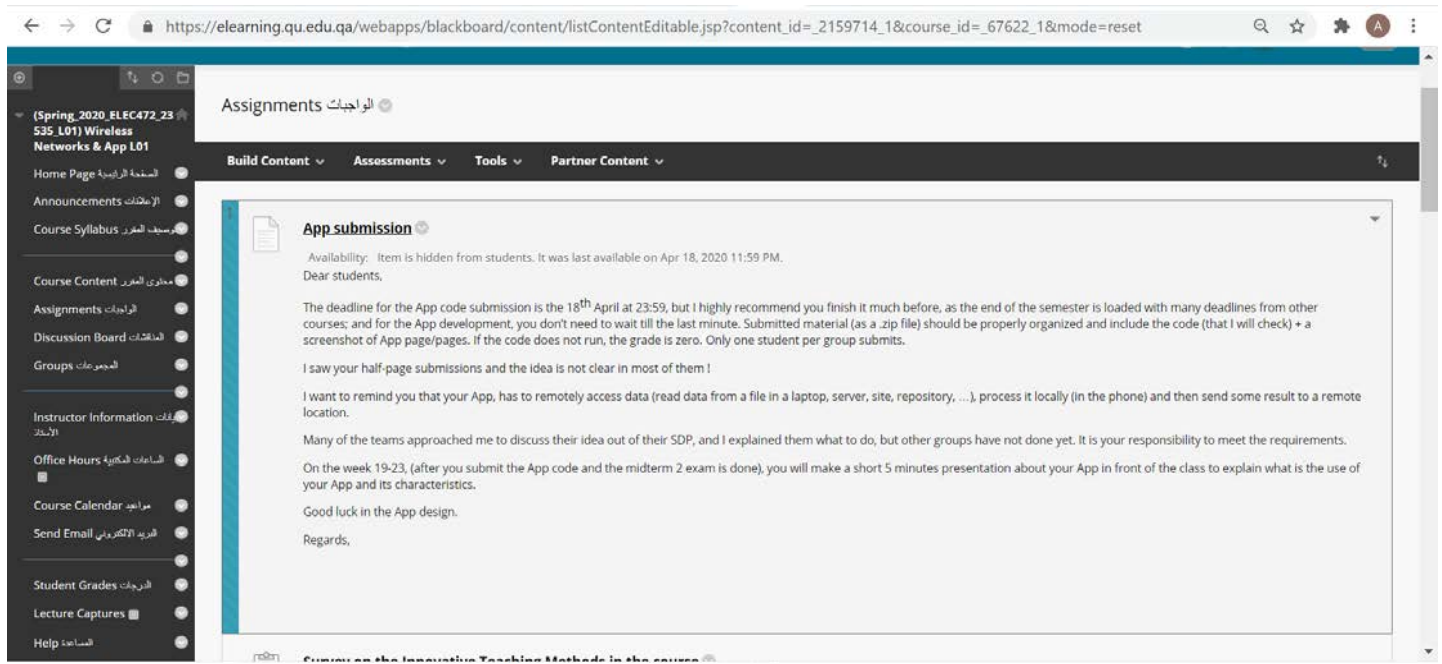

**Figure A1:** Project Guidelines screenshot from the Learning Management System.